\documentclass[aps,showpacs,twocolumn,eqsecnum,floatfix]{revtex4}

\usepackage{epsfig}
\usepackage{latexsym}
\usepackage{mathbbol}

\begin{document}


\title{The cosmic censor conjecture: Is it generically violated?}

\author{Miguel Alcubierre}
\email{malcubi@nuclecu.unam.mx}

\author{Jos\'e A. Gonz\'alez}
\email{cervera@nuclecu.unam.mx}

\author{Marcelo Salgado}
\email{marcelo@nuclecu.unam.mx}

\author{Daniel Sudarsky}
\email{sudarsky@nuclecu.unam.mx}

\affiliation{Instituto de Ciencias Nucleares, Universidad Nacional
Aut\'onoma de M\'exico, A.P. 70-543, M\'exico D.F. 04510, M\'exico.}


\date{\today}


\begin{abstract}
It has been recently argued by Hertog, Horowitz and
Maeda~\cite{Hertog03a}, that generic reasonable initial data in
asymptotically anti deSitter, spherically symmetric, space-times
within an Einstein-Higgs theory, will evolve toward a naked
singularity, in clear violation of the {\it cosmic censor
conjecture}. We will argue that there is a logical and physically
plausible loophole in the argument and that the numerical evidence in
a related problem suggests that this loophole is in fact employed by
physics.
\end{abstract}




\maketitle


\section{Introduction}
\label{sec:introduction}

One of the most famous conjectures of gravitational physics and
perhaps of all physics is the so called {\it cosmic censor conjecture}
(CCC)~\cite{Penrose79}. The physical formulation of this conjecture
states that, for physically reasonable initial data, space-time cannot
evolve toward a naked singularity. That is, if a singularity forms, it
will be covered by an event horizon ({\em i.e.} it will be contained
within a black hole), indicating that far away observers will not be
influenced by it. The importance of the conjecture resides in the fact
that its validity will prevent the complete collapse of the predictive
power of physical laws (a naked singularity will eventually influence
the rest of the space-time), as far as asymptotic observers are
concerned (among which we count ourselves).  On the other hand,
singularity theorems predict the formation of singularities starting
from certain regular initial data when appropriate energy conditions
are satisfied.  However, up to now, an despite the efforts of many
relativists, it has not been possible to prove that such singularities
will always be contained within a black hole (the conjecture resists
to be promoted to the status of a theorem).

Most of the reasonable matter that undergoes gravitational collapse
ends up in the formation of a black hole with a completely regular
domain of outer communication, a fact that produces strong supporting
evidence for the validity of the CCC.  On the other hand, there are
indeed ``counterexamples'' that clearly violate the CCC.  However,
these are not generic in the sense that the initial conditions leading
to the formation of a naked singularity are
fine-tuned~\cite{Choptuik93}. Since it is thought to be physically
impossible to prepare a system with such precise initial conditions,
those counterexamples are considered to be rather artificial, and the
CCC is nowadays considered to apply only in the generic sense.

Recently, however, Hertog, Horowitz and Maeda (HHM)~\cite{Hertog03a},
reported having found a generic counterexample to the CCC. To that
end, they construct an open set of initial data within an
Einstein-Higgs system in asymptotically anti deSitter (AdS),
spherically symmetric, space-times with a scalar-field potential
$V(\phi)$ which is not positive semi-definite.  In this report, we
re-analyze the situation and show that the HHM arguments have a
generic loophole, which we describe in a general setting.
Furthermore, we will argue that although at this point our analysis
does not prove or disprove the example constructed
in~\cite{Hertog03a}, the existing numerical evidence points toward the
realization of our generic loophole rather than the formation of a
naked singularity.


\section{Description}
\label{sec:description}

The situation considered in~\cite{Hertog03a} corresponds to a scalar field
minimally coupled to gravity with Lagrangian
\begin{equation}
{\cal L} = \sqrt{-g} \left[ { 1\over 16\pi }R
- {1\over 2}(\nabla \phi)^2 - V(\phi) \right]  \; .
\end{equation}
(units where $G=c=1$ are employed). 

The scalar field has a tilted Mexican hat potential with a true
minimum at $\phi = \phi_a$ with $V(\phi_a)=-a$, and a local minimum at
$\phi = \phi_b$ with $V(\phi_b)= -b$ ($a>b>0$).  The idea is to
construct initial data that will evolve toward a cosmological type
singularity in a central region, while having a mass that is to small
to allow the formation of a black hole sufficiently large to enclose
the singularity.  To construct this initial data one needs to provide
on a 3-manifold the 3-metric $h_{ab}$, the extrinsic curvature
$K_{ab}$, the scalar field $\phi$ and its time derivative, all
subject to the momentum and hamiltonian constraints. The strategy to
construct the example that is argued to lead to a naked singularity is
as follows:

One considers the manifold $\mathbb{R}^3$ and constructs spherically symmetric
initial data with $K_{ab} = 0$ and $\partial_t\phi = 0$. The tree metric can be
written as
\begin{equation} 
h _{ab} dx^a dx^b = A(r) dr^2 + B(r)r^2 \left( d\theta^2
+ \sin^2\theta d\varphi^2 \right) \; . 
\end{equation}
To find the metric function $A(r)$ one needs only be concerned with
the hamiltonian constraint which, upon the use of the following
re-parametrization
\begin{equation}
A\equiv \left( 1 - \frac{2m(r)}{r}
- \frac{\Lambda_{\rm eff} r^2}{3} \right)^{-1} ,
\end{equation} 
can be cast simply as
\begin{equation}
m'= 4 \pi r^2 \left[ \frac{1}{2A} {\phi'}^2 + V(\phi) + b\right] \; .
\end{equation}
In the case of interest we want to consider a field configuration that
interpolates from a central region with $\phi = \phi_a$ to an exterior
region with $\phi = \phi_b$. We then choose an effective cosmological
constant $\Lambda_{\rm eff}= -8\pi b$, thus ensuring the convergence
of $M = {\rm lim}_{r \to \infty} m(r)$ (provided that asymptotically
$\phi\sim \phi_b + Cr^{-3/2-|\epsilon|}$).  Given the scalar field
configuration we can solve for $m(r)$ from the hamiltonian constraint
above~\cite{Hertog03a} . The initial data is then completely specified
by $\phi(r)$.

One now looks for a class of configurations that would have $\phi (r)
= \phi_a + \epsilon $ for $r<R_0$, and $\phi (r) = \phi_b$ for $r>R_1$
($R_0 < R_1$), with $R_0$ scaling more or less linearly with $R_1$
within the class, and with a total mass that also scales linearly with
$R_1$. The argument would then go as follows: In the domain of
dependence corresponding to the region with $r<R_0$ we will have a
cosmological solution corresponding to anti-De-Sitter space-time with
a scalar field oscillating at the bottom of the potential, which is
known to lead to a singularity within a finite proper time as seen by
co-moving observers. Moreover, we can ensure that there are points on
the initial surface for which all future directed causal curves
emanating from them will hit the singularity.

Now let us assume that the evolution of the initial data produces a
black hole that encloses the singularity. The black hole will
eventually settle to a stationary stage which must correspond to a
Schwarzschild anti-De-Sitter (SAdS) space-time with cosmological
constant corresponding to the asymptotic value of the potential
$\Lambda_{\rm eff}= -8\pi b$, and whose mass can not exceed the mass
available in the initial configuration.  This limits the area of this
late time black hole and, given the fact that the area of the event
horizon is a non-decreasing quantity, it bounds the area of the region
that can be enclosed by any horizon that might be present at the
initial hypersurface. The mass has been arranged to scale with $R_1$
which implies that the bound on the area scales as $R_1^{1/3}$ (at
least for large enough SAdS holes).  For the singularity to be
enclosed by the event horizon, that null surface must intersect either
1) the boundary of the domain of dependence of the initial data region
with $r<R_0$ ({\em i.e.} the in-going null congruence starting at
$r=R_0$ before the generation of the singularity), or 2) the initial
hypersurface at $r>R_0$.  Both these cases would require the area of
the horizon to exceed its bound. This will preclude the singularity
from been enclosed by such horizon, and thus the singularity must be
naked.

The loophole lies in the assumption that the spacetime will evolve
toward a stationary configuration at late times. In fact, it is quite
possible in principle for the following alternative scenario to
develop: The outermost region corresponding to the asymptotic regime
of the SAdS solution with $\Lambda_{\rm eff} = -8\pi b$ is
continuously shrinking (from the inside) due to a domain wall which
expands outward and which corresponds to the scalar field
interpolating between the two minima. The inner region becomes a SAdS
spacetime with a large black hole that encloses the singularity and
which would, if extending infinitely outward, have a mass that exceeds
the initial mass of the space-time, but we must recognize that here we
are dealing with two very different notions of mass corresponding to
two different values of the cosmological constant for the initial and
final AdS space-times one would be considering ($\Lambda^{a}= -8\pi a$
and $\Lambda^{b}= -8\pi b= \Lambda_{\rm eff}$). Of course, in reality
the interior AdS region never covers the whole space-time.  The wall
would behave as a solitonic type solution (such as a boosted ``kink''
in Sine-Gordon system) in that, once initially set it would always
keep moving outward (in our case it could, of course, be changing its
form and needs not be a true soliton).  The apparent paradoxical
situation which would allow the generation of a large black hole which
will be present in any late time hypersurface is explained as due to
the large negative energy which, relative to AdS with cosmological
constant $\Lambda^b$, represents the large inner region with
cosmological constant $\Lambda_a$.


\section{Numerical example}
\label{sec:numerical}

\begin{figure}
\vspace{5mm}
\epsfig{file=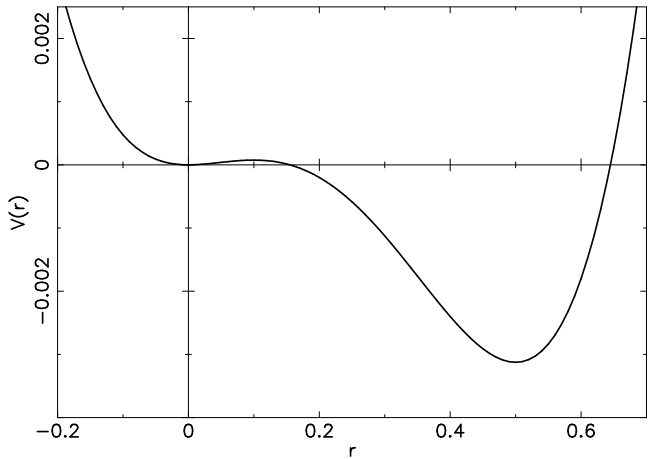,width=6.0cm,angle=270}
\caption{Scalar potential $V(\phi)$ corresponding to
Eq.~(\ref{eq:potential}).}
\label{fig:V} 
\end{figure}

We have performed a numerical simulation of a related system in which
the local minimum of the potential corresponds to zero vacuum energy
($b=0$), which replaces the asymptotic AdS region by an asymptotically
flat region and makes the numerical evolution considerably less
cumbersome. Our potential has the form (see Fig.~\ref{fig:V})
\begin{equation}
V = \frac{1}{4}\phi^2 \left(\phi^2 - \frac{4}{3}(\eta_1 + \eta_2) \phi 
+ 2\eta_1\eta_2 \right) \; ,
\label{eq:potential}
\end{equation}
with $\eta_1 = 0.5$ and $\eta_2 = 0.1$.

For this potential, we have constructed a static unstable soliton as
in Ref.~\cite{Nucamendi03}.  We later perturb this initial
configuration in the form $\partial_t\phi(0,r)= \epsilon e^{-r^2}$
($\epsilon << 1$). We also specify $B(0,r)=1$, and solve the
hamiltonian constraint for $A(0,r)$ and the momentum constraint for
$K_{ab}(0,r)$.  As mentioned, since the static soliton configuration
is unstable, the small perturbation in the scalar field suffices to
trigger dynamical evolution away from the initial configuration. Here
the sign for $\epsilon$ is chosen for the configuration to ``explode''
rather than to collapse to a black hole~\cite{Alcubierre04c}, since we
want to simulate a situation in which the spacetime does not settle to
a stationary state.  The dynamics of the spacetime is followed by
solving the full non-linear 3+1 Einstein evolution
equations~\cite{Arnowitt62,York79} for the spherically symmetric case
(see Ref.~\cite{Alcubierre04c} for details).  For the simulation
considered here we have used Eulerian spatial coordinates ({\em i.e.}
zero shift vector) and harmonic time slicing, which should approach
the interior singularity asymptotically reaching it only after an
infinite coordinate time~\cite{Bona97a,Alcubierre02b} (the lapse
function collapses to zero as the singularity is approached).

\begin{figure}
\vspace{5mm}
\epsfig{file=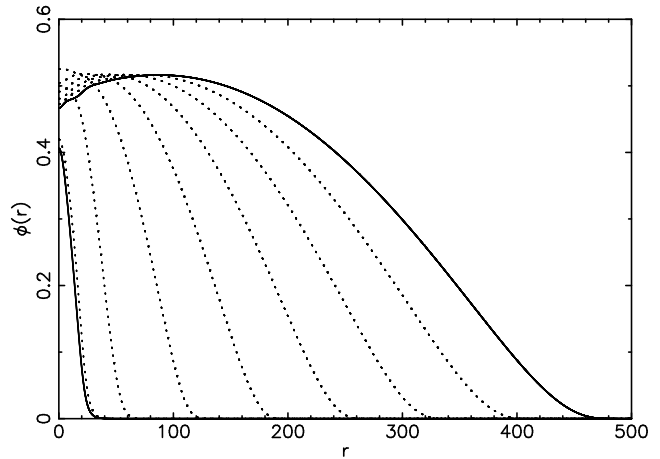,width=6.0cm,angle=270}
\caption{Evolution of the scalar field $\phi$.  Notice how the field
is moving toward the true minimum at $\phi=0.5$ everywhere.}
\label{fig:phi} 
\end{figure}

\begin{figure}
\vspace{5mm}
\epsfig{file=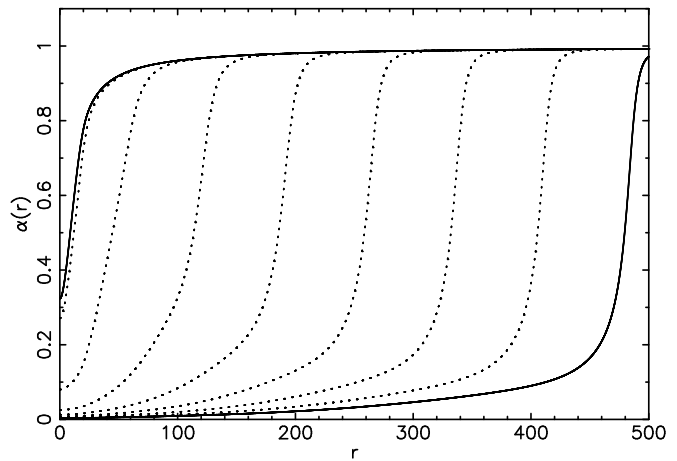,width=6.0cm,angle=270}
\caption{Evolution of the lapse function $\alpha$.  Notice the
collapse of the lapse indicating the approach to a singularity.}
\label{fig:alpha} 
\end{figure}

\begin{figure}
\vspace{5mm}
\epsfig{file=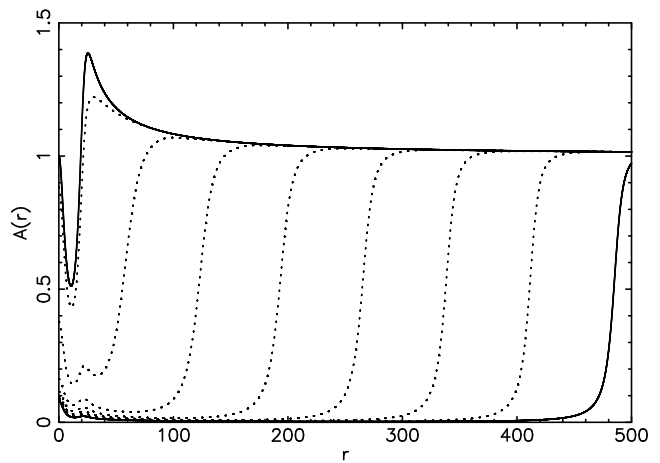,width=6.0cm,angle=270}
\caption{Evolution of the radial metric function A.}
\label{fig:A} 
\end{figure}

The numerical simulation shows that the scalar field tends to move to
the global minimum of the potential everywhere, with a scalar field
``wall'' moving outward.  The result of this is that the outer
Schwarzschild regions (where $V(\phi)\sim 0$) are eventually reached
by the wall of scalar field, becoming inner AdS regions (where
$V(\phi)<0$).  Therefore, the final configuration never reaches a
stationary state.  Figures \ref{fig:phi}-\ref{fig:A} depict this
evolution sequentially from the initial configuration to the point
where the moving wall reaches the outer boundary of the numerical grid
(initial and ``final'' states are shown by solid lines, intermediate
stages by dashed lines).  The numerical evolution eventually freezes
when the lapse function collapses everywhere to zero.  We have looked
for apparent horizons at every time step during the evolution but have
found none.  Notice that the fact that no apparent horizon seems to
form does not imply that no black hole is present, as an event horizon
might very well exist.  The presence of an event horizon seems likely,
as outgoing null lines outside the scalar field wall should reach null
infinity, while those inside should reach the singularity instead.
However, should such an event horizon exist, the black hole will not
reach a stationary state and would eventually swallow the whole
spacetime.

Our simulation shows what will probably happen in the true SAdS
situation. Thus, we conjecture that the corresponding spacetime will
not reach a stationary situation.  We emphasize that the numerical
simulation for SAdS is considerably more difficult than in the
asymptotically flat case for reasons related to numerical stability in
the exterior regions.  However, we are currently working around the
stability problem and will report on this in the near future.

\section{Conclusion}
\label{sec:conclusion}

We conclude that the generic counterexample of the CCC proposed by HHM
need not be such after all.  It is of course still possible that a
naked singularity might arise in their setting, but the question is
fully open and as far as we see the only way to settle this issue is
through a numerical simulation similar to the one we have carried out,
but adapted to the SAdS asymptotics of the HHM example.

The CCC seems to resist attempts in both directions (prove or
disprove) to change its nature of conjecture.


\section{Acknowledgments}
The authors wish to thank G.~Horowitz for useful communications.  This
work was supported in part by \linebreak CONACyT through grant 149945, by
DGAPA-UNAM through grants IN112401 and IN122002, and by DGEP-UNAM
through a complementary grant.


\bibliographystyle{apsrev}
\bibliography{referencias}


\end{document}